\documentclass[prl,aps,epsfig,superscriptaddress,twocolumn]{revtex4-1}
\setcitestyle{super}
\usepackage{amsfonts,amssymb,amscd,amsthm}
\usepackage{graphicx}
\usepackage{mathrsfs}
\usepackage[intlimits]{amsmath}
\usepackage[colorlinks, citecolor=red]{hyperref}
\usepackage{etoolbox}
\usepackage{upgreek}
\usepackage{braket}

\begin{document}
	
\preprint{}

\newcommand{\thetitle}{Observation of quantum superposition of topological defects in a trapped ion quantum simulator}

\author{Z.-J. Cheng}
\thanks{These authors contribute equally to this work}%
\affiliation{Center for Quantum Information, Institute for Interdisciplinary Information Sciences, Tsinghua University, Beijing 100084, PR China}

\author{Y.-K. Wu}
\thanks{These authors contribute equally to this work}%
\affiliation{Center for Quantum Information, Institute for Interdisciplinary Information Sciences, Tsinghua University, Beijing 100084, PR China}
\affiliation{Hefei National Laboratory, Hefei 230088, PR China}

\author{S.-J. Li}
\affiliation{Center for Quantum Information, Institute for Interdisciplinary Information Sciences, Tsinghua University, Beijing 100084, PR China}

\author{Q.-X. Mei}
%\affiliation{Center for Quantum Information, Institute for Interdisciplinary Information Sciences, Tsinghua University, Beijing 100084, PR China}
\affiliation{HYQ Co., Ltd., Beijing 100176, PR China}

\author{B.-W. Li}
%\affiliation{Center for Quantum Information, Institute for Interdisciplinary Information Sciences, Tsinghua University, Beijing 100084, PR China}
\affiliation{HYQ Co., Ltd., Beijing 100176, PR China}

\author{G.-X. Wang}
\affiliation{Center for Quantum Information, Institute for Interdisciplinary Information Sciences, Tsinghua University, Beijing 100084, PR China}
%\affiliation{HYQ Co., Ltd., Beijing 100176, PR China}

\author{Y. Jiang}
\affiliation{Center for Quantum Information, Institute for Interdisciplinary Information Sciences, Tsinghua University, Beijing 100084, PR China}
%\affiliation{HYQ Co., Ltd., Beijing 100176, PR China}

\author{B.-X. Qi}
\affiliation{Center for Quantum Information, Institute for Interdisciplinary Information Sciences, Tsinghua University, Beijing 100084, PR China}

\author{Z.-C. Zhou}
\affiliation{Center for Quantum Information, Institute for Interdisciplinary Information Sciences, Tsinghua University, Beijing 100084, PR China}
\affiliation{Hefei National Laboratory, Hefei 230088, PR China}

\author{P.-Y. Hou}
\affiliation{Center for Quantum Information, Institute for Interdisciplinary Information Sciences, Tsinghua University, Beijing 100084, PR China}
\affiliation{Hefei National Laboratory, Hefei 230088, PR China}

\author{L.-M. Duan}
\email{lmduan@tsinghua.edu.cn}
\affiliation{Center for Quantum Information, Institute for Interdisciplinary Information Sciences, Tsinghua University, Beijing 100084, PR China}
\affiliation{Hefei National Laboratory, Hefei 230088, PR China}
\affiliation{New Cornerstone Science Laboratory, Beijing 100084, PR China}

\title{\thetitle}
\date{\today}
\begin{abstract}
Topological defects are discontinuities of a system protected by global properties, with wide applications in mathematics and physics. While previous experimental studies mostly focused on their classical properties, it has been predicted that topological defects can exhibit quantum superposition. Despite the fundamental interest and potential applications in understanding symmetry-breaking dynamics of quantum phase transitions, its experimental realization still remains a challenge.
Here, we report the observation of quantum superposition of topological defects in a trapped-ion quantum simulator.
By engineering long-range spin-spin interactions, we observe a spin kink splitting into a superposition of kinks at different positions, creating a ``Schrodinger kink'' that manifests non-locality and quantum interference.
Furthermore, by preparing superposition states of neighboring kinks with different phases, we observe the propagation of the wave packet in different directions, thus unambiguously verifying the quantum coherence in the superposition states.
Our work provides useful tools for non-equilibrium dynamics in quantum Kibble-Zurek physics.
\end{abstract}

\maketitle

Topological defects are low-dimensional irregular structures of a system that are protected by the global properties and cannot be created or destroyed by local perturbations \cite{RevModPhys.51.591}. Apart from the fundamental mathematical interest, topological defects also play important roles in various fields in physics such as domain walls in ferromagnetic materials \cite{10.5555/1499044}, vortices in high-temperature superconductors \cite{RevModPhys.66.1125}, magnetic monopoles in grand unified theories \cite{HOOFT1974276}, and cosmic strings in cosmology \cite{Vilenkin:2000jqa}. Topological defects can naturally appear when a system is quickly driven through a continuous phase transition, which is known as the Kibble-Zurek mechanism \cite{kibble1976topology,kibble1980some,zurek1985cosmological,zurek1996cosmological}, because the relaxation time diverges at the transition point in the thermodynamic limit and becomes slower than the quench dynamics at any finite rate.

\begin{figure}[t]
    \centering
    \includegraphics[width=1\linewidth]{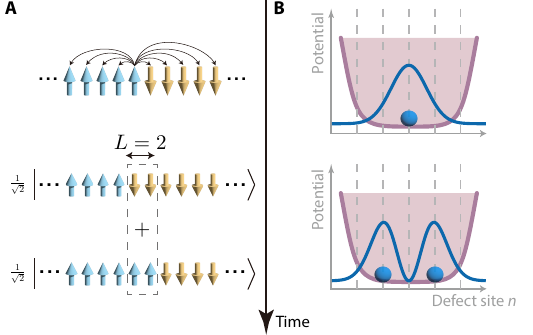}
    \caption{\textbf{Coherent dynamics of topological defects.} \textbf{A}, A spin chain is initialized in a configuration with a topological defect under the ferromagnetic Ising interaction (represented by black arrows in the upper panel). A kink appears in the middle of two ferromagnetic phases with opposite spin orientations and can propagate to the left or the right under a weak transverse field. Consequently, the spin chain evolves into a non-local superposition state, where the kink is simultaneously located at different positions, separated by $L$ sites. \textbf{B}, The dynamics can be understood using a quasi-particle picture. The Ising interaction acts as a potential (purple curve) that confines the kink (blue balls), while the weak transverse field acts as kinetic energy, driving the kink to neighboring sites. As the system evolves into a superposed kink state shown in the lower panel, the wavepackets of the two components interfere and create fringes (blue curve), analogous to the double-slit experiment.}
    \label{fig:1}
\end{figure}

The Kibble-Zurek theory has also been advanced to the quantum regime~\cite{PhysRevLett.95.105701,PhysRevLett.95.035701,PhysRevLett.95.245701,PhysRevB.72.161201} and has recently been demonstrated in the experiment \cite{doi:10.1126/science.aaf9657,anquez2016quantum,keesling2019quantum,PRXQuantum.4.010302}. In particular, it has been predicted that topological defects can be in quantum superposition states \cite{dziarmaga2012non,dziarmaga2022coherent}, with their coherent oscillation governed by the property of the quantum phase transition. However, although quantum superposition has been tested in a wide range of physical systems from elementary particles to mesoscopic and macroscopic objects like ensembles of cold atoms \cite{Kovachy2015}, large organic molecules \cite{Fein2019}, mechanical oscillators \cite{PhysRevLett.130.133604} and superconducting quantum circuits \cite{doi:10.1126/science.1084528}, its demonstration for topological defects remains an experimental challenge. This is because, while topological defects can be viewed as localized quasi-particles, their detection requires measuring global properties of the system and is thus subjected to decoherence and state-preparation-and-measurement (SPAM) errors of the whole system. In previous experiments, only the density of topological defects is concerned, while the phase coherence between different sites has not been certified. Ref.~\cite{dziarmaga2012non} proposed to observe this phase coherence in a spin chain as shown in Fig.~\ref{fig:1}\textbf{A}, where a topological defect (a spin kink) initialized in the middle of the chain can propagate to the left or right. Analogous to a double-slit experiment, the wave packets of the kink moving to the left or the right overlap on the $L$ sites in the middle, resulting in an interference fringe of the probability distribution of the kink location (blue curve in Fig.~\ref{fig:1}\textbf{B}). However, this scheme is still challenging for experiments, as the decoherence rate increases linearly with $L$, bringing the overlapped region into a mixture of two symmetry-breaking states.

Here we report on the observation of quantum superposition of topological defects in a trapped-ion quantum simulator \cite{RMPMonroe}. Among the various physical platforms for quantum simulation, trapped ions have the advantages of high-fidelity state preparation and detection \cite{Quantum-dynamics-of-single-trapped-ions,PhysRevLett.113.220501,PhysRevLett.129.130501}, and long qubit coherence time \cite{Wang2021}. Enabled by individual control of the ion qubits, we initialize the spin chain with a single kink in the middle. We then engineer long-range spin-spin interactions \cite{RMPMonroe} along with a transverse field to coherently split the single kink, allowing us to observe the quantum interference from the superposed spin kink.
Furthermore, we directly prepare the ions in a superposition of two neighboring kinks and observe the directional propagation of the wave packet depending on the relative phase between the two kinks. This unambiguously verifies the quantum coherence in the prepared state. Our work provides useful tools for examining the quantum dynamics of superposed topological defects in quantum Kibble-Zurek physics \cite{dziarmaga2012non,dziarmaga2022coherent}.

\section{Results}
\subsection{Experimental scheme}
We confine a chain of either 20 or 21 $^{171}\rm Yb^{+}$ ions in a blade-style linear Paul trap. The choice of even or odd number of ions is determined by our initial state of a spin kink located at the central one or two sites as described below. The details of the experimental apparatus can be found in our previous works \cite{mei2022experimental,PRXQuantum.4.010302}.
We encode spin states in two clock states, $\ket{\uparrow} \equiv \ket{S_{1/2},F=1,m_F=0}$ and $\ket{\downarrow} \equiv \ket{S_{1/2},F=0,m_F=0}$.
In each experimental trial, ions are first Doppler cooled by near-resonance $370\,$nm laser, then the transverse modes are sideband cooled by counter-propagating global $355\,$nm Raman laser beams to near the ground states \cite{Quantum-dynamics-of-single-trapped-ions,mei2022experimental}. All the ions are prepared in $\ket{\downarrow}$ by optical pumping.
The qubit states are detected simultaneously by using a global resonant laser and an imaging CCD.
The detection error is approximately $2\%$-$3\%$, increasing from central to edge ions.

To observe the dynamics of topological defects, we apply bichromatic Raman laser beams to generate a transverse-field Ising model Hamiltonian \cite{PRXQuantum.4.010302}
\begin{equation}
    \textit{H} = -\sum_{i < j} J_{ij} \sigma_x^i \sigma_x^j + g \sum_i \sigma_z^i \equiv H_{xx} + H_z,
    \label{eq:IsingCoupling}
\end{equation}
where $J_{ij}$ denotes the Ising coupling strength between the $i$-th and $j$-th spins and is governed by the phonon modes of the ion chain and the laser parameters \cite{RMPMonroe}. By setting the laser detuning above the highest transverse phonon mode, we obtain roughly a power-law decay $J_{ij}\approx J_0/|i-j|^\alpha$ with $J_0>0$ and $\alpha \in [1,2]$ typically. The minus sign of the coupling strength is obtained by considering the highest eigenstates of the Ising interaction as the ground states. A uniform transverse field can be generated by shifting the frequencies of the blue- and the red-detuned components of the bichromatic laser in the same direction \cite{mei2022experimental}.

When $g=0$, the two ground states of Eq.~(\ref{eq:IsingCoupling}) are $\ket{+}^{\otimes N}$ and $\ket{-}^{\otimes N}$, where $N$ is the ion number and $\ket{\pm} \equiv(\ket{\uparrow}\pm\ket{\downarrow})/\sqrt{2}$. When these two spin orientations coexist, a domain wall or a spin kink appears at their interface, leading to higher energy. For an Ising model with homogeneous nearest-neighbor interactions, the increased energy will be a constant $2J_0$ independent of the kink location \cite{dziarmaga2012non}. Here due to the long-range interaction, the increased energy $V_n=2\sum_{i\le n,j>n} J_{ij}$ is site-dependent, which can be regarded as a potential for the spin kink defined on a dual lattice with $N-1$ sites, as sketched in Fig.~\ref{fig:1}\textbf{B}.

In the following, we focus on states with a single kink in the form of $\ket{n}\equiv\ket{+}^{\otimes n}\ket{-}^{\otimes (N-n)}$ ($n=1,\cdots,N-1$), which can be prepared by an individual $370\,$nm laser together with global microwave and Raman laser pulses (see Methods). The Ising interaction can be expressed as $H_{xx}^{(1)}=\sum_n V_n \ket{n}\bra{n}$, where the superscript ``$(1)$'' represents the single-kink subspace.
The transverse field $H_z$ can flip all the spins independently, potentially creating more spin kinks with even higher energy. However, in the weak driving regime with $g\ll J_{0}$, such off-resonant excitations are largely suppressed, and we are left with the terms $\sigma_z^n$ and $\sigma_z^{n+1}$ coupling $\ket{n}$ to $\ket{n-1}$ and $\ket{n+1}$, respectively. As a consequence, when restricting to the single-kink subspace, the effect of the transverse field can be approximated by a hopping Hamiltonian
\begin{equation}
H_z^{(1)} =g \sum_{n=1}^{N-2} \ket{n}\bra{n+1}+h.c.   \label{eq:hopping}
\end{equation}

Combining $H_{xx}^{(1)}$ and $H_z^{(1)}$, a spin kink initially localized in the middle of the chain can propagate to the left or right, resulting in an interference pattern in the middle as shown in the lower panel of Fig.~\ref{fig:1}\textbf{B}. Due to the long-range Ising interaction $J_{ij}$ and thus the nonuniform kink potential $V_n$, the wave packet will bounce back before it can propagate far away from the center. Therefore, it is desirable to use a longer ion chain for a wider flat potential in the middle, allowing the spin kink to travel further away and making quantum interference more pronounced. Additionally, using a longer ion chain suppresses edge effects from impacting the dynamics of the kink in the middle. This is because, at the boundaries, the energy required to create a new topological defect is halved, making it more vulnerable to off-resonant excitations. However, increasing the ion number also leads to larger decoherence and SPAM errors, potentially making the observation of quantum interference more difficult. To verify the quantum coherence of the spin kink, it is sufficient to allow the wave packet to propagate by a few sites.

\begin{figure}[t]
    \centering
    \includegraphics[width=1\linewidth]{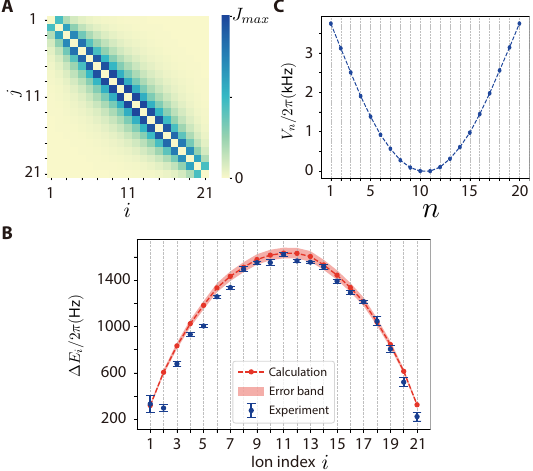}
    \caption{\textbf{Characterization of the Ising interaction and the effective potential of the topological defect.} \textbf{A}, The heatmap of the theoretical spin-spin coupling coefficients $J_{ij}$ in a 21-ion chain, which is calculated from independently calibrated experimental parameters.
    \textbf{B}, Measured spin-flip energies (blue dots) versus the calculated values (red dots) using the $J_{ij}$ matrix in \textbf{A}. Error bars of the experimental data represent one standard deviation. The error band of the theoretical results accounts for a laser frequency fluctuation of $\pm 2\pi \times 500$\,Hz.
    \textbf{C}, The effective potential of the kink calculated from the $J_{ij}$ matrix. We have redefined the zero point of the potential at the center.}
    \label{fig:2}
\end{figure}

\subsection{Calibrating the kink potential}

To calibrate the effective potential $V_n$, we calculate the interaction coefficients $J_{ij}$ using experimental parameters including the carrier Rabi frequency of each ion and the phonon mode parameters, which are calibrated independently in experiments (see Methods).

Specifically, we set the detunings of the bichromatic Raman laser from the carrier transition $\ket{\uparrow}\leftrightarrow \ket{\downarrow}$ to be  $\delta \pm \mu $.
The common detuning $\delta$ of the two tones determines the transverse field strength with $\delta = 2g = 2\pi \times 100$\,Hz.
We set the differential detuning $\mu = \omega_{\rm COM} + 3 \eta_{\rm COM} \Omega_{c}$, where $\Omega_{c}$ is the carrier Rabi frequency of the central ion, $\omega_{\rm COM}=  2\pi \times3.16\,\rm MHz$ and $\eta_{\rm COM}=0.08$ are the frequency and Lamb-Dicke parameter of the center-of-mass (COM) mode, respectively. Such a large detuning from all the transverse motional sidebands ensures that they are only virtually excited with an excitation number bounded by about $0.1$~\cite{RMPMonroe}
The calculated $J_{ij}$ matrix for 21 ions is shown as a heat map in Fig.~\ref{fig:2}\textbf{A}, with $J_{\mathrm{max}}\equiv \max_{ij}\{J_{ij}\}=2\pi\times 184\,$Hz. For the experiment with 20 ions, $J_{\mathrm{max}}=2\pi\times 155\,$Hz is smaller due to the slightly larger ion spacing while the laser intensity remains unchanged.

In principle, the $J_{ij}$ coefficients can also be measured directly in the experiment by the coherent imaging spectroscopy method \cite{CIS}. However, to measure all the $O(N^2)$ coefficients requires $O(N)$ frequency scans and sequentially preparing states with more spin flips. Here for simplicity, we follow our previous work \cite{PRXQuantum.4.010302} to partially verify the calculated $J_{ij}$'s in the experiment. We initialize the spins in $\ket{-}^{\otimes N}$ and measure the spin-flip energy $\Delta E_{i}=2 \sum_{j \ne i} J_{ij}$ for all ions \cite{CIS}. We implement the Hamiltonian $H_{probe} = H_{xx} + B(t) \sum_i \sigma_y^i$ where a weak probe field $B(t) = B_{p} \sin(\omega_p t)$ ($B_p \ll J_0$) is generated by a third tone in the Raman laser resonant with the carrier transition \cite{PRXQuantum.4.010302}. By scanning $\omega_p$ and measuring individual spin states, we obtain $\Delta E_{i}$ for each spin. The experimental results (blue dots in Fig.~\ref{fig:2}\textbf{B}) agree with the theoretical predictions (red dots), in particular for the central ions, indicating the validity of the calculated $J_{ij}$ matrix and the effective potential $V_n$ shown in Fig.~\ref{fig:2}\textbf{C}.

\subsection{Quantum interference of superposed kink states}

\begin{figure}[th]
    \centering
    \includegraphics[width=1\linewidth]{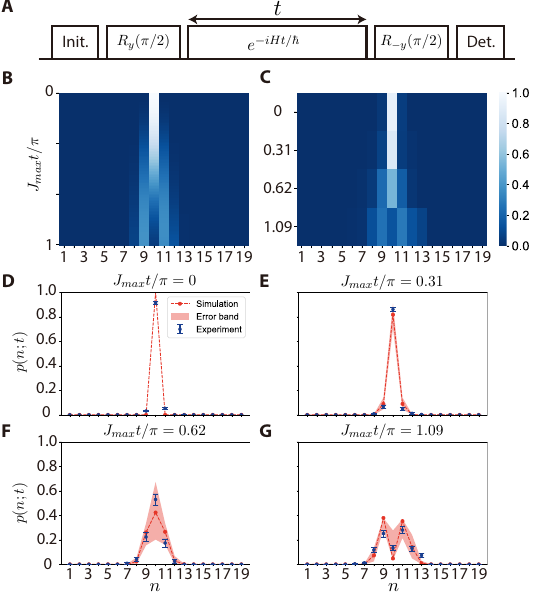}
    \caption{\textbf{Observation of superposed topological defects.} \textbf{A}, Experimental pulse sequence. After initializing the spin chain with a kink in the $\sigma_z$ basis, we apply a global $\pi/2$ pulse to rotate it into the $\sigma_x$ basis. Then the system is evolved for various times $t$ under the transverse-field Ising Hamiltonian. Finally we apply the second $\pi/2$ pulse and the state detection. \textbf{B}, Numerical simulation and \textbf{C}, experimental results of the dynamics of a single kink initialized at the center of a 20-ion chain. We perform measurements at $J_{\mathrm{max}} t/\pi = 0,0.31,0.62,1.09$, with $J_{\mathrm{max}}=2\pi\times 155\,$Hz. \textbf{D}-\textbf{G}, Population distribution of the kink at different sites $p(n)$ for different evolution times. Blue dots are experimental data with one standard deviation error bars. Red dots are simulated data, connected by red dashed lines as the guides to the eyes.  The red error band accounts for a theoretical shift of $\pm 2\pi \times 15\,$Hz in the Raman laser frequency.}
    \label{fig:3}
\end{figure}

We observe the coherent dynamics of a single kink in a 20-ion chain and compare the experimental results to numerical simulations. The experimental pulse sequence is shown in Fig.~\ref{fig:3}\textbf{A}. We initialize the spin chain in $\ket{\uparrow}^{\otimes 10}\ket{\downarrow}^{\otimes 10}$ and then apply a global Raman laser $\pi/2$ pulse to rotate it into $\ket{n=10} = \ket{+}^{\otimes 10}\ket{-}^{\otimes 10}$.
The system is evolved under the Hamiltonian~(\ref{eq:IsingCoupling}) for various times $t$. Finally, we rotate the spins back to the $\sigma_z$ basis for single-qubit detection.

Ideally, the number of topological defects is conserved during the evolution. However, due to experimental imperfections and decoherence, we observe that more than one kink can appear and that the average number of kinks grows as the evolution time increases. We post-select measured spin configurations with a single kink and normalize the obtained probability distribution $p(n;t)$ ($n=1,\,2,\,...,\,19$) of the kink being at the $n$-th site after an evolution time of $t$.

The heatmap for the measured probabilities $p(n;t)$ is shown in Fig.~\ref{fig:3}\textbf{C} and is in good agreement with the numerical simulation based on the effective single-kink Hamiltonian $H_{xx}^{(1)}+H_z^{(1)}$ as displayed in Fig.~\ref{fig:3}\textbf{B}. In particular, an interference pattern is observed as the wave packet propagates to the left and right.
The dynamics can be seen more clearly from the data measured at distinct time points in Figs.~\ref{fig:3}\textbf{D}-\textbf{G} with $J_{\mathrm{max}}t/\pi=0,\, 0.31,\, 0.62,\, 1.09$.
The experimental data (blue dots) are plotted together with the simulation results (red dots) and an error band (red shaded regions) which represents a frequency drift of $\Delta g =\pm 2\pi \times 15\,$Hz for the Raman lasers (our scheme is less sensitive to frequency drifts in the phonon frequency \cite{mei2022experimental}).
When $t=0$ (Fig.~\ref{fig:3}\textbf{D}), we obtain $p(n;t=0)\approx \delta_{n,10}$ as expected. Small populations at neighboring sites are primarily due to imperfections in state preparation.
As the evolution time $t$ grows, the wave packet spreads to the left and the right, and we observe a broader distribution $p(n;t)$ in  Figs.~\ref{fig:3}\textbf{E}, \textbf{F}.
However, distinct from the classical diffusion of a wave packet (see Supplementary Notes for more details), here the quantum wave packet can demonstrate an interference pattern from the superposition of the left- and right-going components.
In Fig.~\ref{fig:3}\textbf{G} with $J_{\mathrm{max}}t=1.09\pi$, we clearly see two peaks in $p(n;t)$ at $n=9$ and $n=11$, with a suppressed probability at $n=10$ due to the destructive interference. For this $N=20$ case, the two wave packets will start to bounce back at a longer evolution time. For a longer ion chain, theoretically the wave packet can propagate farther away from the center and it is possible to observe more interference fringes.
%However, this also means larger decoherence, which is dominated by the Raman laser dephasing, and thus lower data rate after post-selection.

\subsection{Directional propagation of superposed kink states}

\begin{figure}
    \centering
    \includegraphics[width=0.9\linewidth]{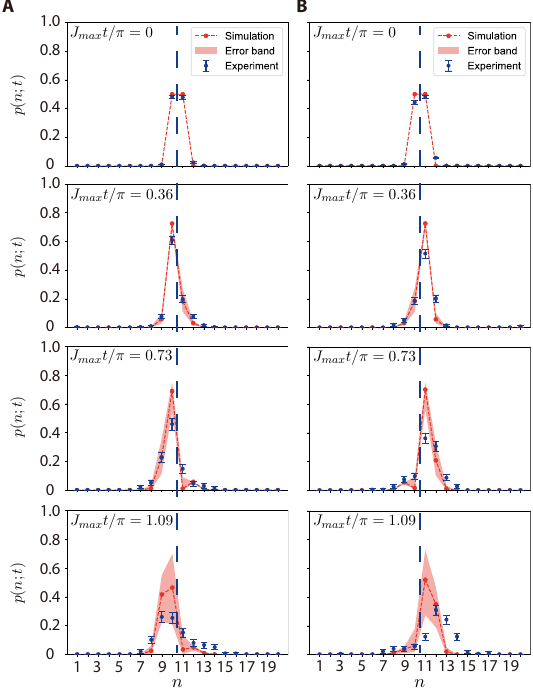}
    \caption{\textbf{Directional propagation of a superposed topological defect.} Time evolution of an initial superposed kink with a relative phase \textbf{A}, $\pi/2$ or \textbf{B}, $-\pi/2$ at various times $t$.
    Blue dots are experimental data with one standard deviation error bars. Red dots are simulated data connected by red dashed lines as the guides to the eyes, while the red error band accounts for a shift of $\pm 2\pi \times 15$Hz in the Raman laser frequency.
    Whether the superposed kink propagates to the left or right depends on the initial phase of the superposition state, verifying its quantum coherence. The vertical dashed lines denote the initial location of the kink.
    }
    \label{fig:4}
\end{figure}

Finally, we directly verify the quantum coherence of the superposed kink by showing that the different initial phases in the superposition states can lead to the propagation of the kink in different directions.
For this purpose, we consider a chain of $N=21$ ions with $N-1=20$ sites for the spin kink. The experimental sequence is similar to that in Fig.~\ref{fig:3}\textbf{A}, while the ions are prepared in a superposition state $(\ket{n=10}\pm i\ket{n=11})/\sqrt{2}$ before the Ising evolution (see Methods). According to Eq.~(\ref{eq:hopping}), after a short evolution time $\Delta t$, each localized kink can propagate to the neighboring sites with a relative amplitude of $-ig\Delta t$. Therefore, when we initialize the kink at two adjacent sites with a relative phase of $\pm \pi/2$, it will result in constructive interference in one direction and destructive interference in the other direction, hence the directional propagation of the wave packet.

Similar to Fig.~\ref{fig:3}, we measure the kink population $p(n;t)$ at $J_{\mathrm{max}}t/\pi=0,\, 0.36,\, 0.73,\, 1.09$ in Fig.~\ref{fig:4}.
Again we compare the experimental results (blue dots with error bars) with numerical simulations (red dots with error bands).
We observe that the wave packet of the single kink moves to the left for an initial phase of $\pi/2$ (Fig.~\ref{fig:4}\textbf{A}) while to the right with a phase of $-\pi/2$ (Fig.~\ref{fig:4}\textbf{B}).
We note that the experimental results for the initial phase of $-\pi/2$ show larger deviations from the simulation results compared to the $\pi/2$ phase because of larger crosstalk errors in state preparation (see Methods).

\section{Conclusion}
% what we did
In summary, we observe the non-equilibrium quantum dynamics of a single spin kink in a one-dimensional ion chain. We verify the quantum coherence of a superposed spin kink by observing the quantum interference pattern and the directional propagation controlled by the initial phase.
% Discuss limitation and how to improve?
More intriguing phenomena may be observed by preparing more spin kinks \cite{PhysRevB.102.014308,bennewitz2024simulating}, including more ions, and increasing the evolution time. The last two are currently limited by the laser intensity and the coherence time of our apparatus, which will be improved in future upgrades.
% what's next and what are the potential implications/impact.
The approach may also be extended to the recently developed two-dimensional trapped-ion simulator \cite{guo2024site} for exploring coherent dynamics of other types of topological defects.

\bigskip

\textbf{Data Availability:} The data that support the findings of this study are available
from the authors upon request.

\textbf{Acknowledgements:} This work was supported by Innovation Program for Quantum Science and Technology (2021ZD0301601), Tsinghua University Initiative Scientific Research Program, and the Ministry of Education of China. L.M.D. acknowledges in addition support from the New Cornerstone Science Foundation through the New Cornerstone Investigator Program. Y.K.W. acknowledges in addition support from Tsinghua University Dushi program. P.Y.H. acknowledges the start-up fund from Tsinghua University.

\textbf{Competing interests:} Q.X.M. and B.W.L. are affiliated with HYQ Co. Y.K.W., Q.X.M., B.W.L., Y.J., B.X.Q., Z.C.Z. and L.M.D. hold shares with HYQ Co. The other authors declare no competing interests.

\textbf{Author Information:} Correspondence and requests for materials should be addressed to L.M.D.
(lmduan@tsinghua.edu.cn).

\textbf{Author Contributions:} L.M.D. proposed and supervised the project. Z.J.C., P.Y.H., S.J.L., Q.X.M., B.W.L., G.X.W, Y.J., B.X.Q., Z.C.Z. carried out the experiment. Y.K.W. did the associated theory and data analyses. Y.K.W., Z.J.C., P.Y.H., and L.M.D. wrote the manuscript.

%\bibliographystyle{naturemag_nourl}
%\bibliography{references}

\appendix
\makeatletter
\renewcommand{\thefigure}{{\color{black}Extended Data Fig.~}\arabic{figure}}
\renewcommand{\thetable}{{\color{black}Extended Data Table~}\arabic{table}}
\renewcommand{\theHfigure}{Extended Data Fig.~\arabic{figure}}
\renewcommand{\theHtable}{Extended Data Table~\arabic{table}}
\renewcommand{\fnum@figure}{\thefigure}
\renewcommand{\fnum@table}{\thetable}
\setcounter{figure}{0}
\setcounter{table}{0}
\makeatother
\makeatletter
\apptocmd{\thebibliography}{\global\c@NAT@ctr 33\relax}{}{}
\makeatother

\section{Methods}
\subsection{Topological-defect state preparation}
\begin{figure}
    \centering
    \includegraphics[width=1\linewidth]{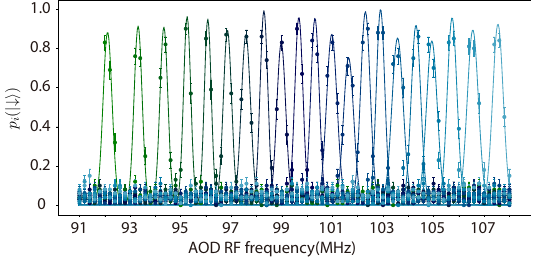}
    \caption{\textbf{Individual addressing of 21 ions using an acousto-optic deflector.} A 370\,nm laser pulse embedded in a Ramsey experiment individually addresses a single ion, causing a local AC Stark shift. The $i$-th ion has a higher probability $p_i(\ket{\downarrow})$ ending up in $\ket{\downarrow}$ if the individual beam is applied onto this ion.
    The individual beam location is controlled by the RF frequency of the AOD. By scanning the RF frequency, we calibrate the required frequencies for addressing each ion indicated by the resonances in the plot. Curves in different colors represent the results of different ions.}
    \label{fig:S1}
\end{figure}

We prepare initial topological defect states by using a Ramsey interferometry sequence, where an individual addressing laser pulse is embedded to cause AC Stark shift on selected ions \cite{PhysRevA.94.042308}. The individual addressing laser beam is at $370\,$nm and detuned approximately $43.6\,$GHz from the $S_{1/2}\rightarrow P_{1/2}$ transition.
This beam addresses one ion at a time and the beam pointing is controlled by the RF frequency of an acousto-optic deflector (AOD). Figure~\ref{fig:S1} shows the RF frequencies to address each ion in the 21-ion chain.
The pulse duration to realize an $R_z(\pi)$ rotation is further calibrated for each ion.
This individual addressing beam has a half width at half maximum of approximately $1.3\,\mu$m compared to the ion spacing ranging from $4.7$ to $8.0\,\mu$m. This can cause crosstalk errors, defined as the probability of changing the qubit state of a spectator ion during individual addressing operations. Crosstalk errors are more pronounced when addressing central ions compared to edge ions. Due to beam aberration, the crosstalk error is asymmetric and is measured to be about $2.5\%$ for the left side and about $5\%$ for the right side when addressing central ions. In the experiment, we choose to address the ions on the right side to minimize state preparation errors caused by crosstalk. The crosstalk error to the left side when addressing the first target ion is smaller, while the crosstalk error to the right side can be compensated by adjusting the duration of the next pulse and so on.

To prepare the ion chain with a kink in the middle of the 20-ion chain, we first apply a global microwave $\pi/2$ pulse followed by the individual addressing pulses to implement the $R_z(\pi)$ rotation for the right ten ions. The second microwave $\pi/2$ pulse brings the right ten ions to $\ket{\downarrow}$ while the left ten ions to in $\ket{\uparrow}$. A Raman laser $\pi/2$ pulse can be applied to prepare the ions in the desired single-kink state $\ket{n=10}\equiv\ket{+}^{\otimes 10}\ket{-}^{\otimes 10}$ while maintaining the phase coherence with the $\sigma_x$ basis of the Ising interaction. To mitigate the error due to the inhomogeneous laser intensity, we use the composite BB1 pulse \cite{CompositePulse} for the global $\pi/2$ pulses.

To prepare a 21-ion chain in a superposition state with two neighboring kinks, $(\ket{n=10} \pm i \ket{n=11})/ \sqrt{2} = \pm i\ket{+}^{\otimes 10} \ket{\mp y} \ket{-}^{\otimes 10}$, where $\ket{\pm y} = (\ket{\uparrow} \pm i \ket{\downarrow})/ \sqrt{2}$. We first prepare the ions in $\ket{+}^{\otimes 10}\ket{-}^{\otimes 11}$, similarly to the 20-ion chain preparation.
We further apply an additional $R_z(\pi/2)$ or $R_z(3\pi/2)$ rotation on the $11$-th ion using the individual addressing beam, thus preparing the $11$-th ion in $\ket{\pm y}$.
Note that the $R_z(3\pi/2)$ pulse is longer and causes larger crosstalk errors on nearby ions than the $R_z(\pi/2)$ rotation, which is reflected on the experimental results shown in Fig.~\ref{fig:4}.

\subsection{Calculation of the effective spin-spin interaction}
\begin{figure*}[th]
    \centering
    \includegraphics[width=1\linewidth]{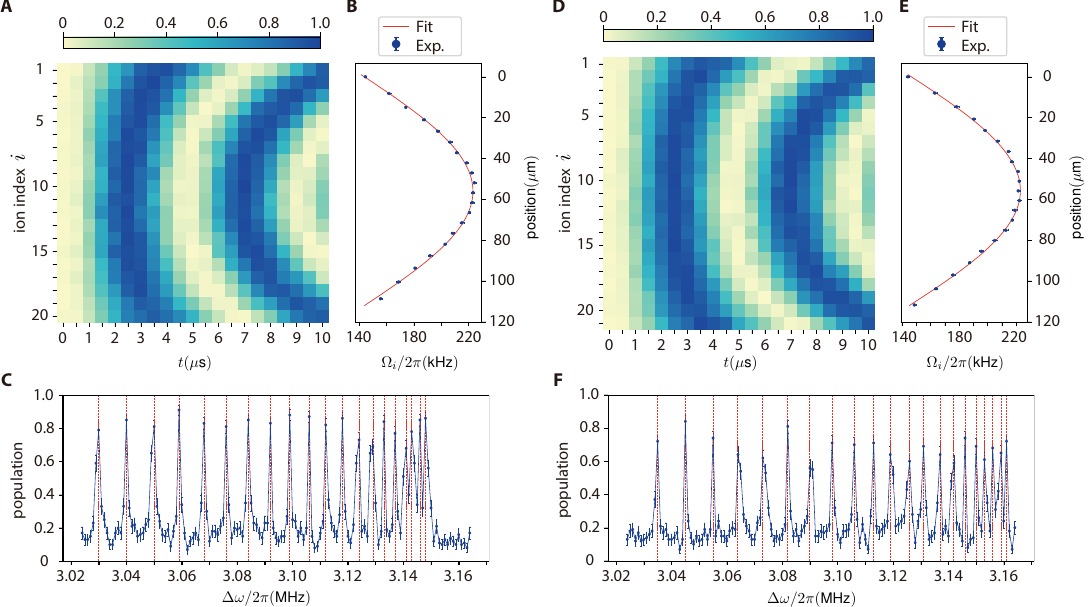}
    \caption{\textbf{Calibration of carrier Rabi frequencies and phonon mode frequencies.}
    Panels \textbf{A}-\textbf{C} and panels \textbf{D}-\textbf{F} display the calibration results for a 20-ion chain and a 21-ion chain, respectively.
    \textbf{A}, \textbf{D}, Heatmaps of carrier transition probabilities for all ions as a function of Raman laser pulse duration.
    The Rabi oscillations of individual ions are separately fitted to a theory model, yielding their Rabi frequencies, which are shown as blue dots in \textbf{B} and \textbf{E}. The vertical axis represents ions' position with the origin being the position of the first ion. The measured Rabi frequencies versus the ion position follow a Gaussian distribution whose full width at half maximum is fitted to be around $143\,\mu$m. Fits are shown as red solid lines. \textbf{C}, \textbf{F},
    Blue-sideband spectra (blue dots), which are the transition probability as a function of the detuning $\Delta\omega$ from the carrier transition, are used to extract the phonon mode frequencies (red dashed lines). All data points are with 68\% confidence error bars.}
    \label{fig:S2}
\end{figure*}

The Ising coupling coefficients are calculated by using $J_{ij} = \Omega_i \Omega_j \sum_k \eta_k^2 b_{ik} b_{jk} \omega_k / (\mu^2 - \omega_k^2)$ from the experimental parameters \cite{RMPMonroe,PRXQuantum.4.010302}.
The carrier Rabi frequencies $\Omega_{i}$ of each ion under the global Raman laser beams are obtained by preparing all ions in the $\ket{\downarrow}$ state, driving the carrier transition by the Raman lasers, and measuring the qubit states after various evolution times $t$. The experiments are repeated for 200 times, and the average transition probability of each ion is calculated and shown as heatmaps in \ref{fig:S2}\textbf{A} for a 20-ion chain and in \ref{fig:S2}\textbf{D} for a 21-ion chain.
The Rabi oscillation for each ion is fitted to yield the Rabi frequency $\Omega_i$, which is represented by blue dots in \ref{fig:S2}\textbf{B} and \ref{fig:S2}\textbf{E}. The Rabi frequencies follow a Gaussian distribution (red curves) with the full width at half maximum fitted to be approximately $143\,\mu$m.

The phonon mode frequencies $\omega_k$ are measured from the blue-sideband spectrum as shown in \ref{fig:S2}\textbf{C} and \ref{fig:S2}\textbf{F} for the two ion crystals, respectively. We first cool all the transverse modes near the ground states by sideband cooling and initialize all the ions in $\ket{\downarrow}$. Then we apply a weak global Raman pulse near the blue motional sideband at a variable frequency with a fixed duration, and finally we perform the fluorescence detection. In this calibration, the scattered photons are counted by a photomultiplier. The photon counts of 100 experimental trials are compared to a threshold of one to obtain the probability that at least one ion has been successfully excited. The probabilities are plotted versus the detuning from the carrier transition $\Delta \omega$ in \ref{fig:S2}\textbf{C} and \ref{fig:S2}\textbf{F} for a 20-ion and a 21-ion chain, respectively.

These measured phonon mode frequencies, together with a rough estimation of the ion spacings from the image of the ion chain on the CCD camera, are then used to fit more accurate ion spacings \cite{mei2022experimental}. These data further allow us to calculate the normal mode vectors $b_{ik}$, the Lamb-Dicke parameters $\eta_k$, and finally the $J_{ij}$ matrix shown in Fig.~\ref{fig:2}\textbf{A}.

\subsection{Numerical simulation in single-kink subspace}
The numerical simulations in the main text are performed for a quasi-particle restricted in the single-kink subspace under the Hamiltonian
\begin{equation}
H^{(1)} = \sum_{n=1}^{N-1} V_n \ket{n}\bra{n} + g\sum_{n=1}^{N-2} (\ket{n}\bra{n+1} + \ket{n+1}\bra{n}),
\end{equation}
where the effective potential is given by $V_n=2\sum_{i\le n,j>n} J_{ij}$. This is a quantum system with a total dimension of $N-1$, and can be computed efficiently even for a large $N$.

It is worthwhile to compute the energy gap between the single-kink subspace and the other multi-kink states that can be reached by the Hamiltonian evolution. Specifically, here we consider the excitation energy when applying $\sigma_z^i$ on $\ket{n}$ with $i\ne n,n+1$, which is given by
\begin{equation}
\Delta E(n,i) = \left\{
\begin{array}{ll}
2(\sum\limits_{j=1}^{i-1}+\sum\limits_{j=i+1}^{n}-\sum\limits_{j=n+1}^{N})J_{ij} & (i<n) \\
2(\sum\limits_{j=n+1}^{i-1}+\sum\limits_{j=i+1}^{N}-\sum\limits_{j=1}^{n})J_{ij} & (i>n+1)
\end{array}
\right.
\end{equation}
For simplicity, here we use the power-law fitting $J_{ij}\approx J_0/|i-j|^\alpha$ with $\alpha\approx 1.3$. As expected, such energies are smallest when the kink is near the boundary ($n\approx 0$ or $n\approx N-1$), but for $n\approx N/2$ as in our experiment, the energy gap is larger than $3J_0$. Therefore, by setting $g=J_0/3=2\pi\times 50 \rm Hz$ in this experiment, the probability of escaping from the single-kink subspace due to the off-resonant excitation will be small. In the experiment, we attribute the main error sources for the leakage outside the single-kink subspace to the decoherence of the system.

\end{document}